\documentclass[12pt,a4paper]{article}

\usepackage{epsf}
\usepackage{amsmath}
\usepackage{amssymb}
\usepackage{bbm}
\usepackage{rotate}
\usepackage[hypertex]{hyperref}
\renewcommand{\baselinestretch}{1.5}
\makeatletter \@addtoreset{equation}{section} \makeatother

\newcommand{\be}{\begin{equation}}
\newcommand{\ee}{\end{equation}}
\newcommand{\bea}{\begin{eqnarray}}
\newcommand{\eea}{\end{eqnarray}}
\newcommand{\op}{\xi}

\newcommand{\cL}{\mathcal{L}}

\newcommand{\pV}{\mathbbm{V}}

\newcommand{\rmd}{\mbox{\rm{d}}}
\newcommand{\e}{\mbox{\rm{e}}}

\newcommand{\qd}{\dot{q}}
\newcommand{\pd}{\dot{\phi}}
\newcommand{\ad}{\dot{\alpha}}
\newcommand{\bd}{\dot{\beta}}

\newcommand{\ft}[2]{{\textstyle\frac{#1}{#2}}}
\begin{document}
\begin{titlepage}
\begin{center}
\renewcommand{\baselinestretch}{1} \large\normalsize
\hfill hep-th/0410273 \\
\hfill FSU-TPI-08/04 \\
\hfill ITP-UU-04/37 \\
\hfill SPIN-04/20 \\
\renewcommand{\baselinestretch}{1.5} \large\normalsize
\renewcommand{\thefootnote}{\fnsymbol{footnote}}
\vskip 0.7cm {\large \bf  Dynamical Conifold Transitions and Moduli Trapping  \\
in M-Theory Cosmology}\footnote{Work supported by the `Schwerpunktprogramm Stringtheorie' of the DFG.}
\renewcommand{\thefootnote}{\arabic{footnote}}
\setcounter{footnote}{0}

\vskip 0.7cm

{\bf Thomas Mohaupt$^a$ and Frank Saueressig$^b$}  \\[3ex]
\renewcommand{\baselinestretch}{1} \large\normalsize
$^a${\em Institute of Theoretical Physics,
Friedrich-Schiller-University 
Jena, \\ 
 Max-Wien-Platz 1, D-07743 Jena, Germany} \\
{\tt Thomas.Mohaupt@uni-jena.de}\\[2.5ex]
$^b${\em Institute of Theoretical Physics} and {\em Spinoza Institute, \\
Utrecht University, 3508 TD Utrecht, The Netherlands} \\
{\tt F.S.Saueressig@phys.uu.nl} \\ 
\renewcommand{\baselinestretch}{1.5} \large\normalsize
\end{center}
\vskip 0.7cm
\begin{center} {\bf ABSTRACT } \end{center}
\vspace{-3mm}
We study five-dimensional Kasner cosmologies in the vicinity of a conifold locus occurring in a time-dependent Calabi-Yau compactification of M-theory. The dynamics of M2-brane winding modes, which become light in this region, is taken into account using a suitable gauged supergravity action. 
We find cosmological solutions which interpolate between the two branches of the transition, establishing that conifold transitions can be realized 
dynamically.
However, generic solutions do not correspond to transitions, but to the moduli
getting trapped close to the conifold locus. This effect results from an 
interplay between the scalar potential and Hubble friction. We show
that the dynamics does not depend on the details of the potential, but
only on its overall shape.
\noindent 
\end{titlepage}


\begin{section}{Introduction}

Topology change 
\cite{AspGreMor,GreMorStr,Witten:1993yc,Witten:1996qb}
is one of the most remarkable properties of string and M-theory 
compactifications on special holonomy manifolds.\footnote{We refer
to  \cite{Green_review,Acharya:2004qe} for review and more
references.} 
While such changes are, by definition, discontinuous processes from the
viewpoint of differential geometry, they are smooth within string theory in
the sense that all observables are continuous. The usual low energy effective
action (LEEA) which includes the generically massless modes of the
compactification only, however, becomes discontinuous or even singular at the
transition locus. It was the insight of \cite{Str1} that these discontinuities
or singularities can be attributed to string or brane winding modes which
become massless in the transition. Following \cite{beauty} we will
call points in the moduli space where additional massless states 
occur
`extra species points' (ESPs). Topological phase transitions then correspond 
to a subclass of ESPs. In these cases the internal space develops a 
singularity which can be smoothed in more than one way, thereby connecting 
internal spaces with different topologies. 

So far, topological phase transitions have mostly been discussed in terms
of parametric deformations of the internal manifold. 
The moduli encoding
 these deformations appear as scalars in the LEEA and parameterize the ground
 state of the action. In the vicinity of a topological phase transition (or, 
more generally, any ESP) one obtains additional light modes which should be
explicitly included in the LEEA. A program to find such `extended' LEEA for 
topological phase transitions (and other types of ESPs) 
has been started in \cite{MohProc,SU2,LMZ} and was recently extended to flop \cite{FS3} and conifold \cite{CFL,DIS} transitions occurring in Calabi-Yau (CY) compactifications of M-Theory. In the latter cases the extra light modes correspond to charged matter multiplets. These originate from M2-branes wrapping 
the internal two-cycles which shrink to zero volume at the transition locus. Including these extra states in the extended LEEA induces a scalar potential which is uniquely determined by the geometry of the transition. 
This potential does not arise from switching
on fluxes: it is a consequence of the
presence of charged matter and supersymmetry.
Therefore it is intrinsic to flop and conifold transitions.  

Comparing to flops, the special 
feature of a conifold transition is the 
appearance of a new branch of the moduli space, a so-called Higgs-branch, which starts at the
transition point. In terms of the LEEA the conifold 
transition point is the intersection point between the Coulomb and the Higgs branch
of an Abelian gauge theory. Geometrically, the transition between these two
branches corresponds to a topological phase transition in which the Hodge
numbers (and therefore the Euler number)
of the internal space changes.\footnote{We refer to \cite{Green_review} 
for a review of the geometrical aspects.} 

Concerning the dynamics of the moduli in the vicinity of a topological phase transition only little is known, however.
For M-theory compactifications on CY threefolds explicit solutions of the
dimensionally reduced theory which interpolate through a flop transition
either as a function of a space-like coordinate or of time have been found in
\cite{GMMS,GSS} and \cite{flop,FS4}, respectively. In the latter case it was
thereby found that the moduli dynamically stabilize in the vicinity of the
flop region. These results complement the ones of
\cite{Berliner,Kachru,Taylor:1999ii} who find that when switching on background fluxes, the remaining 
vacua tend to be located at ESPs.

In this paper we use a consistent truncation of the effective supergravity description for conifold transitions \cite{CFL,DIS} to study the dynamics of the scalar fields in the transition region.
The reason why we work
in five dimensions is that in this case the actions for conifold transitions are 
easier to construct, because the vector multiplet sector of the theory
can be treated exactly \cite{CFL,DIS}. 
The generalization
to four-dimensional conifold LEEA can be treated along the lines
of \cite{LMZ}. We do not expect that the qualitative behavior
of four-dimensional models will be different from the one
found in this paper.

Based on the LEEA \cite{CFL,DIS} we find that the leading order contribution
of the scalar potential is (up to the different numerical factor) given by the
potential $\pV = \frac{1}{2} x^2 y^2$  discussed in \cite{Helling1,beauty}. In
order to make contact to the analytic results obtained there, we 
study the dynamics of the scalar fields in both the full supergravity model and its (non-supersymmetric) leading order approximation. Within the full model we thereby obtain
numerical solutions which 
interpolate between the Coulomb and the Higgs branch of the conifold transition.  But we also
observe that due to Hubble friction such transitions tend
to be suppressed, while the trapping of the moduli in the transition 
region is favored. 
When comparing these results to the leading order approximation it turns out that the qualitative 
behavior of the solutions is the same, and therefore only depends
on the shape of the potential, but not on its details.
This is interesting in the context 
of vacuum selection, because it shows that regions of the moduli space where additional light states appear are dynamically preferred. Investigating the origin of this trapping mechanism we find that it results from an interplay between the shape of the scalar potential and Hubble friction. 

These results (and those of \cite{flop,FS3}) are closely related
though somewhat complementary to those of \cite{beauty}. There the 
mechanism which traps the scalar fields at ESPs is `quantum moduli trapping,'
which is a consequence of the quantum production of light particles occurring when the 
moduli pass near an ESP \cite{beauty}.
In contrast, the trapping mechanism discussed in 
our paper (and in \cite{flop,FS3}) results from an interplay between
the scalar potential and Hubble friction. This has been called
`classical moduli trapping' in \cite{beauty}. In \cite{beauty} these two trapping 
mechanisms have been compared using the results of \cite{Helling1}. We re-investigate this comparison based on our results and find that, once Hubble friction is taken
into account, classical moduli trapping is much more efficient
than found in \cite{beauty}.

\end{section}

\begin{section}{The models}
Our starting point is a five-dimensional Lagrangian of the form
\be\label{2.1}
\sqrt{-g}^{-1} \cL   =   - \frac{1}{2} R  
- \frac{1}{2} g_{XY} \partial_{\mu} q^X \partial^{\mu} q^Y 
- \frac{1}{2} g_{xy} \partial_{\mu} \phi^x \partial^{\mu} \phi^y  
- \pV(\phi, q) \, . 
\ee
Here $g$ is the determinant of the space time metric, $R$ is its Ricci scalar,
$\phi^x$ and $q^X$ are scalar fields of a non-linear sigma model, taking
values in target manifolds with metrics $g_{xy}$ and $g_{XY}$, respectively, and  
$\pV(\phi, q)$ is the scalar potential. Both the (truncated) supergravity description of a conifold transition \cite{CFL,DIS} and the five-dimensional version of the 
model \cite{Helling1,beauty} are of this general form. 

In the LEEA for M-theory compactified on a Calabi-Yau threefold,
the scalars $\phi^x$ and $q^X$ sit in vector and hypermultiplets,
respectively. 
In this case the indices take values $x=1,\ldots, n_V$ and $X=1,\ldots, 4n_H$, where
$n_V$ is the number of vector and $n_H$ the number of hypermultiplets.
The full supergravity Lagrangian for conifold transitions also contains
fermions and gauge fields, and some of the hypermultiplet scalars 
(those which come from wrapped M2-branes) are charged. In order to have a tractable problem, we consider the 
minimal model for a conifold transition \cite{CFL,DIS}, containing one vector
and two hypermultiplets. As explained in \cite{CFL,DIS} it is 
currently not possible to derive the metric on the hypermultiplet
manifold from M-theory. However, the non-compact Wolf spaces
$U(n_H,2)/(U(n_H) \times U(2))$ provide a consistent choice and then the
M-theory charges carried by the wrapped M2-branes fix the
LEEA uniquely.

Restricting the hypermultiplet scalars to their real part and setting the
vector fields to zero provides a consistent truncation of the full
equations of motion.\footnote{This is analogous to the consistent
truncation used for dynamical flop transitions, which is 
described in more detail in \cite{FS3,FS4,DIS}.}
Additionally taking\footnote{Here $v^1, u_1$ and $v^2, u_2$ denote the complex
  hypermultiplet scalars of the first and second hypermultiplet, 
respectively. See \cite{CFL,DIS} for the details.}
\be
{\rm Re}(u_1) = {\rm Re}(u_2) =: q \; , \quad {\rm Re}(v_1) = {\rm Re}(v_2) = 0 \, ,
\ee
the dynamics of the remaining scalar fields 
can be derived from the Lagrangian (\ref{2.1}) by substituting
the scalar field metrics
\be\label{7.8}
g_{xy}  =   \frac{3 \left( 2 + \phi^2 \right)}{(2 - 3 \phi^2)^2 } \; , \qquad 
g_{XY}  =  \frac{2}{\left( 1 - 2 q^2 \right)^2} \; , 
\ee
and the scalar potential
\be \label{7.9}
\pV(\phi, q) =  \left( 48 \pi \right)^{2/3} \, \frac{\phi^2 \, q^2}{\left( 1 - 2 q^2 \right) \, \left(1 - \frac{3}{2} \phi^2 \right)^{2/3}} \; ,
\ee
into the equations of motion given below. 

The scalar potential is positive semi-definite
and vanishes along the lines $q = 0$ (the Coulomb branch)
and $\phi = 0$ (the Higgs branch). The absolute values  of $\phi$ and $q$ can be used as order parameter for the conifold transition.
The vacuum structure of the full supergravity Lagrangian was analyzed in
\cite{CFL,DIS}. Along the Coulomb branch, $q=0$, which is parameterized
by $\phi$, the vector multiplet is massless, while the two 
hypermultiplets, which are charged under the $U(1)$,
have masses proportional to $|\phi|/(2-3 \phi^2)^{1/3}$, the volume of the wrapped cycle. At the 
point $\phi=q=0$, which corresponds to a Calabi-Yau space with 
a conifold singularity, all three multiplets are massless. At this
point one can
give a vev $q\not=0$ to the hypermultiplets, while freezing $\phi=0$. 
This results in one hypermultiplet and the 
vector multiplet combining into a massive long vector multiplet, i.e., a Higgsed $U(1)$, 
while one hypermultiplet remains massless. The points on the Coulomb and Higgs branch (away from the origin) thereby
correspond to two families of smooth Calabi-Yau spaces with different
Hodge numbers, which are related by the conifold transition.
Specifically, we have $\tilde{h}^{1,1} = h^{1,1} -1$, 
$\tilde{h}^{1,2} = h^{1,2} + 1$, where $h^{p,q}$ and 
$\tilde{h}^{p,q}$  are the Hodge numbers of the smooth Calabi-Yau
space corresponding to the Coulomb and Higgs branch, respectively.
The Euler numbers are related by $\tilde{\chi} = \chi -4$.

The scalar metrics $g_{xy}, g_{XY}$ as well as the scalar
potential $\pV$ become infinite for $|\phi| = \sqrt{\ft23}$ and
$|q| = \sqrt{\ft12}$. Since these points are at infinite geodesic distance
from any point with $|\phi| <  \sqrt{\ft23}$ and
$|q| < \sqrt{\ft12}$, we see that the scalar manifold has the 
topology of an open disc and carries a metric such that the boundary is
at infinite distance. Moreover the metric is diagonal, so that
 the scalar manifolds can be
mapped isometrically to 
$\mathbbm{R}^2$ with its standard flat metric. 
In order to compare with the model discussed 
in \cite{Helling1,beauty}, it is convenient to perform this map
explicitly. By solving the geodesic equations along the Coulomb and
the Higgs branch, one finds new coordinates $x,y$ which are equal
to the geodesic lengths: 
\be\label{7.10}
x = \frac{1}{\sqrt{3}} \, \mbox{arctanh} \left( \frac{\phi \, ( \phi^2 - 6 )}{(\phi^2 + 2)^{3/2}} \right) \; , \qquad y =  \mbox{arctanh}(  \sqrt{2} \, q ) \, .
\ee
The Lagrangian now takes the form:
\be \label{7.7}
\sqrt{-g}^{-1}  {\cal L} = - \ft12 R - \ft12 \partial_\mu x \partial^\mu x
- \ft12 \partial_\mu y \partial^\mu y - \pV(x,y)
\ee
where
\be
\label{7.7a}
\pV(x,y) =  \frac{1}{3} \, \, (48 \pi)^{2/3} x^2 y^2 + \mbox{higher order terms} \;.
\ee
With respect to these coordinates the scalar potential takes a more complicated form than in 
(\ref{7.9}), but it is clear that it starts with a term 
proportional to $x^2 y^2$ and diverges for $|x|,|y| \rightarrow \infty$.
Up to the prefactor, the leading order potential is then the same as studied in \cite{Helling1}, while \cite{beauty} discusses a variant where one of the
scalars is complex.\footnote{Thus we cover the situation
where the `impact parameter' $\mu$ of \cite{beauty} (which is the imaginary
part of the complex field) is set to zero.}

When studying the dynamics arising from the Lagrangian (\ref{2.1}), we use the Kasner ansatz 
\be \label{7.1}
ds^2_5 = -  \rmd \tau^2 + \e^{2 \alpha(\tau)} \rmd \vec{x}^2 + \e^{2 \beta(\tau)} \rmd y^2  \, ,
\ee
for the five-dimensional space-time metric. Here $\vec{x} = (x^1, x^2, x^3 )$ are three space-like coordinates, parameterizing the macroscopic dimensions, while $y$ is the coordinate of the fifth, extra dimension. Taking the scalar fields $\phi^x$ and $q^X$ to be homogeneous, this ansatz gives rise to Friedmann's equation
\be \label{7.3a}
3 \left( \ad^2 + \ad \, \bd \right)  =  T +  \pV \, , 
\ee
and the equations of motion
\bea \label{7.3}
2 \ddot{\alpha} + \ddot{\beta} + 2 \ad \bd + 3 \ad^2 + \bd^2  &=& - T + \pV \, , \\
\nonumber 3 \left( \ddot{\alpha} + 2 \ad^2 \right) & = & - T +  \pV \, ,
\eea
in the gravitational and
\bea \label{7.4}
\ddot{\phi}^x + \gamma^x_{~yz} \, \pd^y \, \pd^z + \left( 3 \ad + \bd  \right) \pd^x +  g^{xy} \frac{\partial \pV}{\partial \phi^y } & = & 
0 \, , \\
%
%
\label{7.5}
\ddot{q}^X + \Gamma^X_{~YZ} \, \qd^Y \, \qd^Z + \left( 3 \ad + \bd  \right) \qd^X +  g^{XY} \frac{\partial \pV}{\partial q^Y } & = & 0 \, , 
\eea
in the matter sector, respectively. Here the ``over-dot'' indicates a derivative with respect to the cosmological time $\tau$ and $T$ is
the positive semi-definite kinetic energy 
\be \label{7.6}
T := \frac{1}{2} \, g_{XY} \, \qd^X \, \qd^Y + \frac{1}{2} \, g_{xy} \, \pd^x \pd^y \, .
\ee
Note that the scalar equations of motion are geodesic equations which
are modified by a friction term and a force term. The friction term,
the Hubble friction, comes from 
the coupling to gravity, while the force term reflects the
existence of a scalar potential, which 
couples the vector and hypermultiplet scalars.
While in absence of Hubble friction the energy $T+\pV$ of the
scalar fields is conserved, $T+\pV$ decreases in an expanding
universe ($3 \dot{\alpha} + \dot{\beta} > 0$) and increases
in a contracting universe ($3 \dot{\alpha} + \dot{\beta} < 0$).

\end{section}

\begin{section}{Dynamical conifold transitions}
We will now use the supergravity model specified by the eqs.\ (\ref{7.8}) and
(\ref{7.9}) to give an explicit example of a cosmological solution 
which undergoes a dynamical conifold transition. In order to judge whether a solution evolves along the Higgs or the Coulomb branch we introduce the order parameter
\be
\op := \frac{|q|}{|\phi|} \, ,
\ee
and adopt the criterion that for $\op \gg 1$ ($\op \ge 100$, say) the solution
runs along the Higgs branch while for $\op \ll 1$ ($\op \le 1/100$, say) it
evolves along the Coulomb branch.\footnote{As it will turn out below, it is difficult to find an intrinsic criterion for when a transition is completed. The definition given above provides a reasonable choice for our setup, while in the context of the full string theory setup one might prefer a different definition (see the discussion below).}

We then take the initial conditions
\be\label{3.1}
\begin{split}
\phi(0) = & 0 \, , \quad \dot{\phi}(0) = 0.09 \, , \quad q(0) = 0.65 \, , \quad \dot{q}(0) = 0 \, , \\ 
\quad \alpha(0) = &  0 \, , \quad \dot{\alpha}(0) = 0.1 \, , \quad \beta(0) = 0 \, ,
\end{split}
\ee
with $\dot{\beta}(0)$ being determined by Friedmann's equation, and solve the corresponding equations of motion numerically.

The left and right diagram of figure \ref{vierzehn} display the evolution of
the scalar fields $\phi(\tau), q(\tau)$ and of the logarithmic scale factors $\alpha(\tau), \beta(\tau)$, respectively. 
\begin{figure}[t]
\renewcommand{\baselinestretch}{1}
\epsfxsize=0.4\textwidth
\begin{center}
\leavevmode
\epsffile{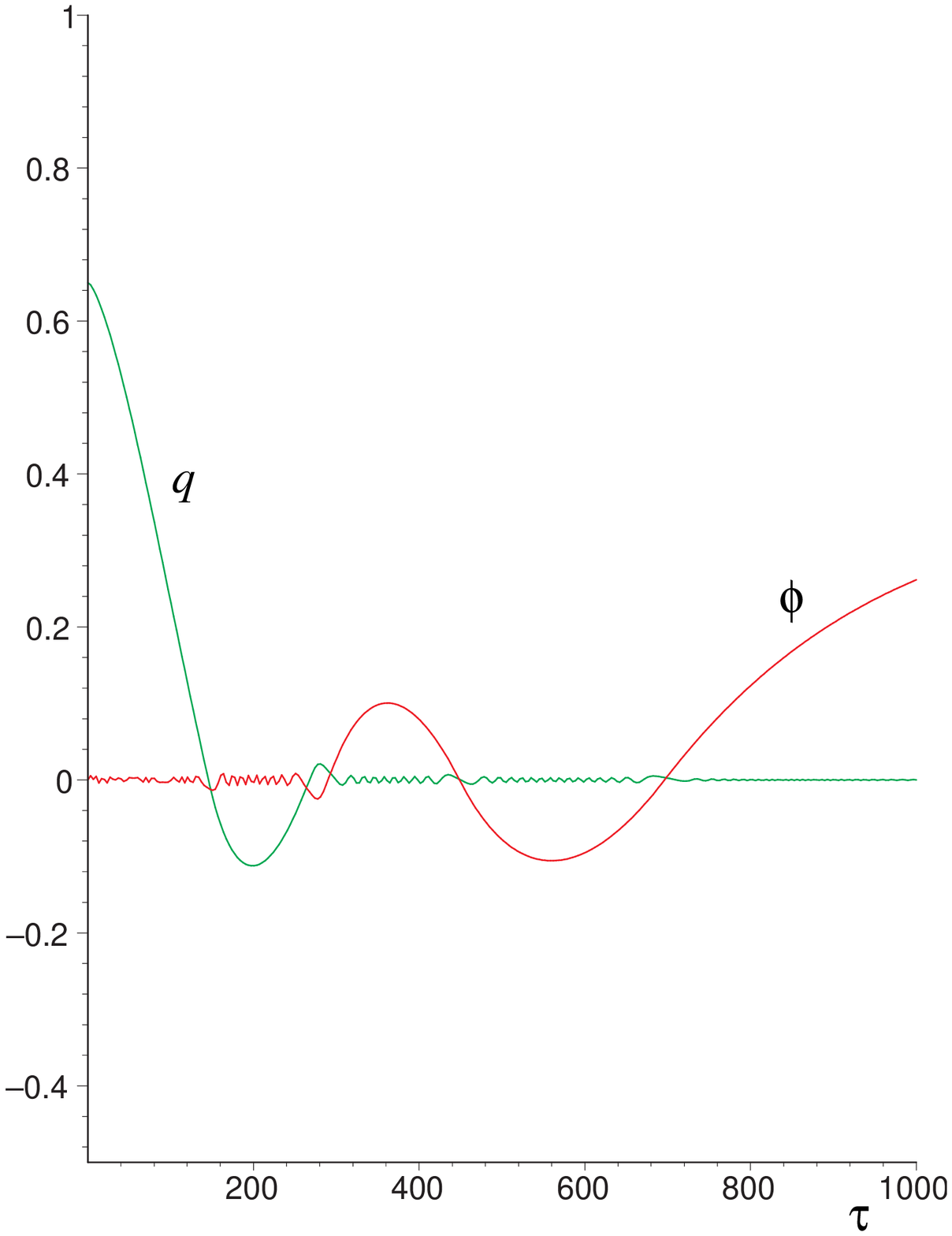} \, \, \, \, \, \,\,
\epsfxsize=0.4\textwidth
\epsffile{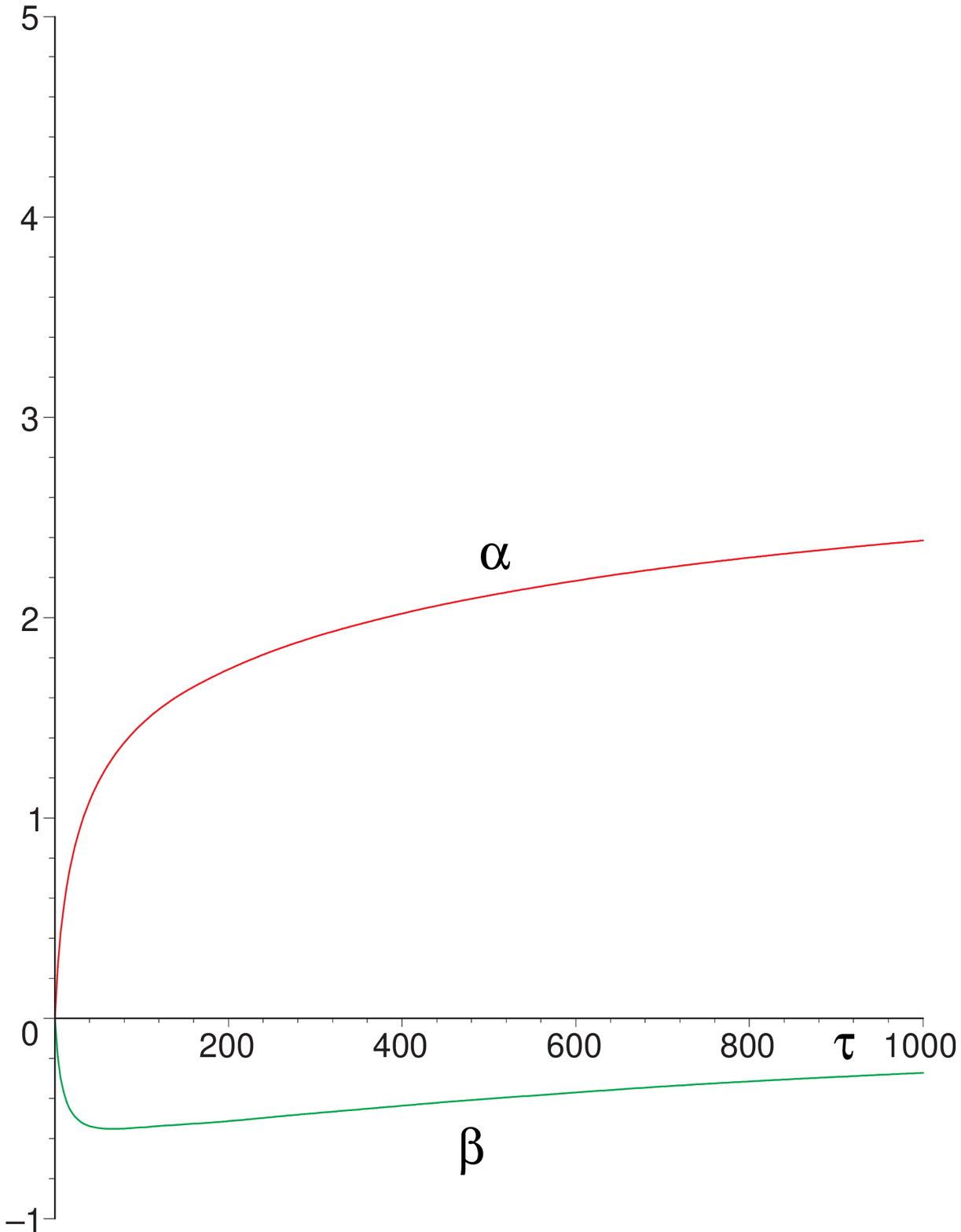}
\end{center}
\parbox[c]{\textwidth}{\caption{\label{vierzehn}{\footnotesize The scalars and logarithmic scale factors along a Kasner cosmological solution undergoing a conifold transition. The initial conditions are given in eq.\ (\ref{3.1}).}}}
\end{figure}
The logarithmic scale factors thereby exhibit an initial period from $0 \le \tau \lesssim 200$, say, with a rapid increase of $\alpha(\tau)$ and decreasing $\beta(\tau)$. During this period the four-dimensional universe expands by a factor of five. After this initial period for large $\tau$ the both of the scale factors are almost constant.

The interesting feature of this solution is the dynamics of the scalar
fields. For $\tau \lesssim 250$, say, the absolute value of $q(\tau)$ is large
while the one for $\phi(\tau)$ is very small ($\op \gg 1$), indicating that the solution
evolves along the Higgs branch. For $250 \lesssim \tau \lesssim 350$ we
encounter a crossover, where $|q(\tau)|$ and $|\phi(\tau)|$ are comparable and
the solution evolves in the central region. For $\tau \gtrsim 350$ the roles
of $q(\tau)$ and $\phi(\tau)$ are reversed. Now $|q(\tau)|$ is small 
while $|\phi(\tau)|$ has become large $(\op \ll 1)$. This indicates that the solution now evolves along the Coulomb branch, revealing that it has undergone a transition from one vacuum branch to the other.

We remark that a conifold transition requires non-vanishing initial values for
$\phi$ or $\dot{\phi}$, and $q$ or $\dot{q}$ as both $\phi = \dot{\phi} = 0$
and $q = \dot{q} = 0$ are consistent solutions of the equations of motion. 
Solutions with initial conditions $\phi_{\rm init} = \dot{\phi}_{\rm init} = 0$
always stay on the Higgs branch while those with 
$q_{\rm init} = \dot{q}_{\rm init} = 0$ remain on the Coulomb branch. 
This implies that conifold transitions can only be described with an 
extended LEEA which explicitly 
includes all the massless states appearing at the conifold point, since otherwise we have $q_{\rm init} = \dot{q}_{\rm init} = 0$ or $\phi_{\rm init} = \dot{\phi}_{\rm init} = 0$ automatically. 

Figure \ref{funfzehn} shows the solution (\ref{3.1}) projected onto the
$\phi$-$q$-plane. The black lines, which close to the origin have the 
shape of a hyperbola,  
illustrate the equipotential lines of the scalar potential. They seem to 
converge at the points $(|\phi|=\sqrt{\ft23}, q=0)$ and $(\phi=0, |q|=
\sqrt{\ft12})$, but this is an artefact of our presentation, because these
points are at infinite geodesic distance, as discussed in the 
previous section.
The Coulomb and the Higgs branch are given by the vertical $q=0$ and the
horizontal $\phi =0$ axes, respectively. The conifold transition corresponds
to the solution `bending around the corner' and going from 
one `valley' to the other. From this picture it is also clear that
one needs initial values where the scalar fields either start  
a little `uphill', or have a non-vanishing
velocity in the `uphill' direction, in order to be able to cross from one branch to the other.

The other remarkable feature of figure \ref{funfzehn} is that it
shows the existence of an effective repulsive force which
drives the scalar fields from the valleys to the central `stadium.'
Observe that the solution turns back twice from an excursion into
a valley, even though its motion is roughly aligned  to the flat direction.
Also note that in its third move into a valley, which we interpret
as a complete transition, it does not move as deeply into the
Coulomb branch as it started in the Higgs branch.  For the potential
$\pV = \ft12 x^2 y^2$ which is, up to the different prefactor, the leading term of the conifold potential (\ref{7.10}), it was shown 
analytically in \cite{Helling1} that there is an effective
potential for the motion along the valleys, which drives solutions
towards the stadium. This is intuitively plausible, because away
from the stadium the valleys become very narrow, 
resulting in a repulsive force acting on solutions which are not perfectly
aligned to the flat directions. For the motion along the $x$-direction, say,
the effective potential found in \cite{Helling1} is 
$\pV_{\rm eff} \propto \ln(x)$. In the conifold case there will be
corrections, but the qualitative properties are the same. 
However, as we will discuss in the next section, this effective
potential is only half of the story. The second effect, which 
strongly enhances the dynamical preference of solutions to stay in
the stadium is the Hubble friction resulting from the coupling to
gravity.
\begin{figure}[t]
\renewcommand{\baselinestretch}{1}
\epsfxsize=0.4\textwidth
\begin{center}
\leavevmode
\epsffile{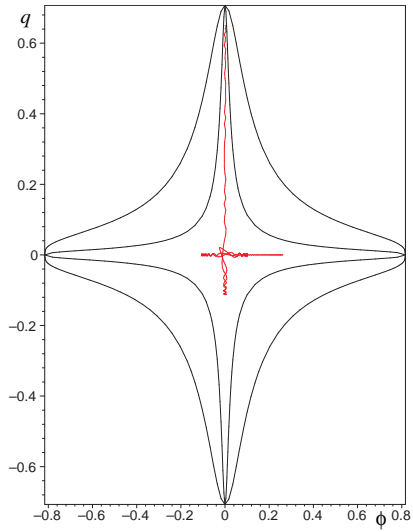} \, \, \, \, \,
\end{center}
\parbox[c]{\textwidth}{\caption{\label{funfzehn}{\footnotesize The cosmological solution (\ref{3.1}) projected to the $q$-$\phi$-plane. The black lines are equipotential lines of the scalar potential. The four corners of the cross correspond to the boundaries of the scalar manifolds which are at infinite geodesic distance from the origin. The conifold transition reflects itself in the solution bending around the corner, switching from the Higgs to the Coulomb branch.}}}
\end{figure}

In our model generic solutions will 
ultimately be driven back to the stadium, an exception being those which are
perfectly aligned to one of the flat directions as these run along their
respective branch for infinite cosmological time. This is a limitation
of our effective supergravity approach to topological phase transitions,
which can only be overcome by lifting the discussion to the full
string theory. Then the relevant scalar manifold does not just
have two branches, but is the huge web (or `stratification') 
composed of all the moduli spaces of Calabi-Yau spaces which can 
be connected by topological transitions. In fact, it is widely 
believed that the moduli space of all Calabi-Yau spaces form just
one single connected web \cite{Reid}. This web has many nodes, each of which 
corresponds to a (higher-dimensional version of a) stadium of the type
considered in this paper. From this perspective, a 
plausible criterion for a completed
transition between two branches is to require that a solution
moves away from the stadium it has crossed by a distance which
is larger than the distance to the next stadium. Clearly, 
the quantitative formulation and test of such a criterion would require
to know the true metric on the web of moduli spaces. Since this is a 
formidable problem,
we leave this challenge for future investigation and use the criterion 
introduced above.

Our numerical studies also yield some evidence that the motion of scalars in 
the conifold potential is chaotic since evolving two solutions with almost
identical initial conditions may result in the two solutions evolving at two
different vacuum branches at some later point in time. This is similar to
observations made for the potential (\ref{7.7}) which is also believed to
exhibit chaotic behavior. Note that the valley structure of our
potential provides one of the ingredients for chaos, namely a sequence of 
bifurcations. We, however, lack another typical ingredient,
namely a hyperbolic fixed point of the equations of motion with both 
attractive and repulsive directions, so that the bifurcations 
lead to a stretching and folding of phase space volumes. 
As in flop transitions \cite{FS3}, the equations of motion have a
family of non-hyperbolic fixed points, given by the vacuum manifolds $q = 0$ or $\phi = 0$, respectively. Therefore the standard methods of the theory
of dynamical systems do not apply, and we fall short of a complete
proof of chaotic behavior.
\end{section}

\begin{section}{Moduli stabilization via Hubble friction}
In the last section we noted that the dynamics of scalar fields
showed a preference for the central region around the conifold point,
or stadium. We now investigate this effect
in detail. By comparing solutions with different, randomly chosen
initial conditions we realize that for generic solutions the scalar
fields get trapped in the stadium. Below we will give a 
representative example. In order to check to which extent the effect
depends on the details of the potential, we present solutions for
both the full conifold model (\ref{7.8}), (\ref{7.9}) and its leading
order approximation (\ref{7.7}) which (up to a constant factor) 
has been studied in \cite{Helling1,beauty}. In order
to compare the two cases we will transform the conifold model into the 
$(x,y)$-basis using eq.\ (\ref{7.10}).
We will also explore the role of Hubble friction by contrasting
the numerical solutions of the full scalar equations of motion
(\ref{7.4}), (\ref{7.5})  to solutions where the Hubble friction term 
 proportional to $3 \ad + \bd$ has been switched off. 
All together we have the following four cases: \\
\hspace*{3mm} i)   leading order approximation without Hubble friction, \\
\hspace*{3mm} ii)  leading order approximation with Hubble friction, \\
\hspace*{3mm} iii) conifold model without Hubble friction, \\
\hspace*{3mm} iv)  conifold model with Hubble friction. \\
In all cases we display solutions for the following initial
conditions:
\be\label{3.2}
\begin{split}
 \phi(0) = & \, 0.05 \, , \quad \dot{\phi}(0) = 0 \, , \quad q(0) = 0.6 \, , \quad \dot{q}(0) = 0 \\ 
\Longrightarrow & \; \; x(0) =  -0.061 \, , \quad \dot{x}(0) = 0 \, , \quad y(0) = 1.251 \, , \quad \dot{y}(0) = 0 \, ,  \\ 
\quad \alpha(0) = &  \, 0 \, , \quad \dot{\alpha}(0) = 0.1 \, , \quad \beta(0) = 0 \, . 
\end{split}
\ee
%
The initial value of $\dot{\beta}$ is again obtained from the condition that the
initial values need to satisfy Friedmann's equation. 
\begin{figure}[p!]
\renewcommand{\baselinestretch}{1}
\epsfxsize=0.4\textwidth
\begin{center}
\leavevmode
\epsffile{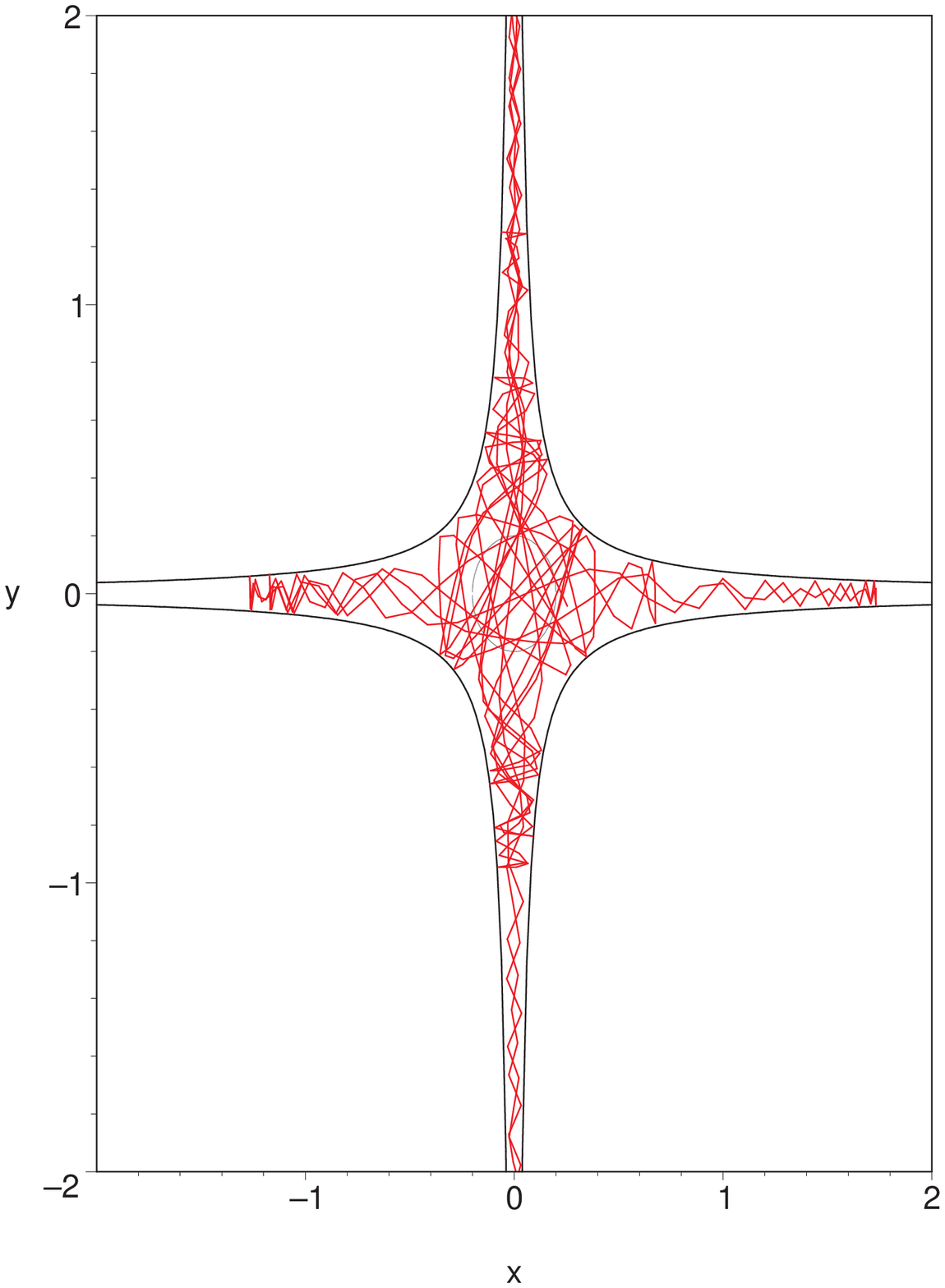} \,  
\epsfxsize=0.4\textwidth
\epsffile{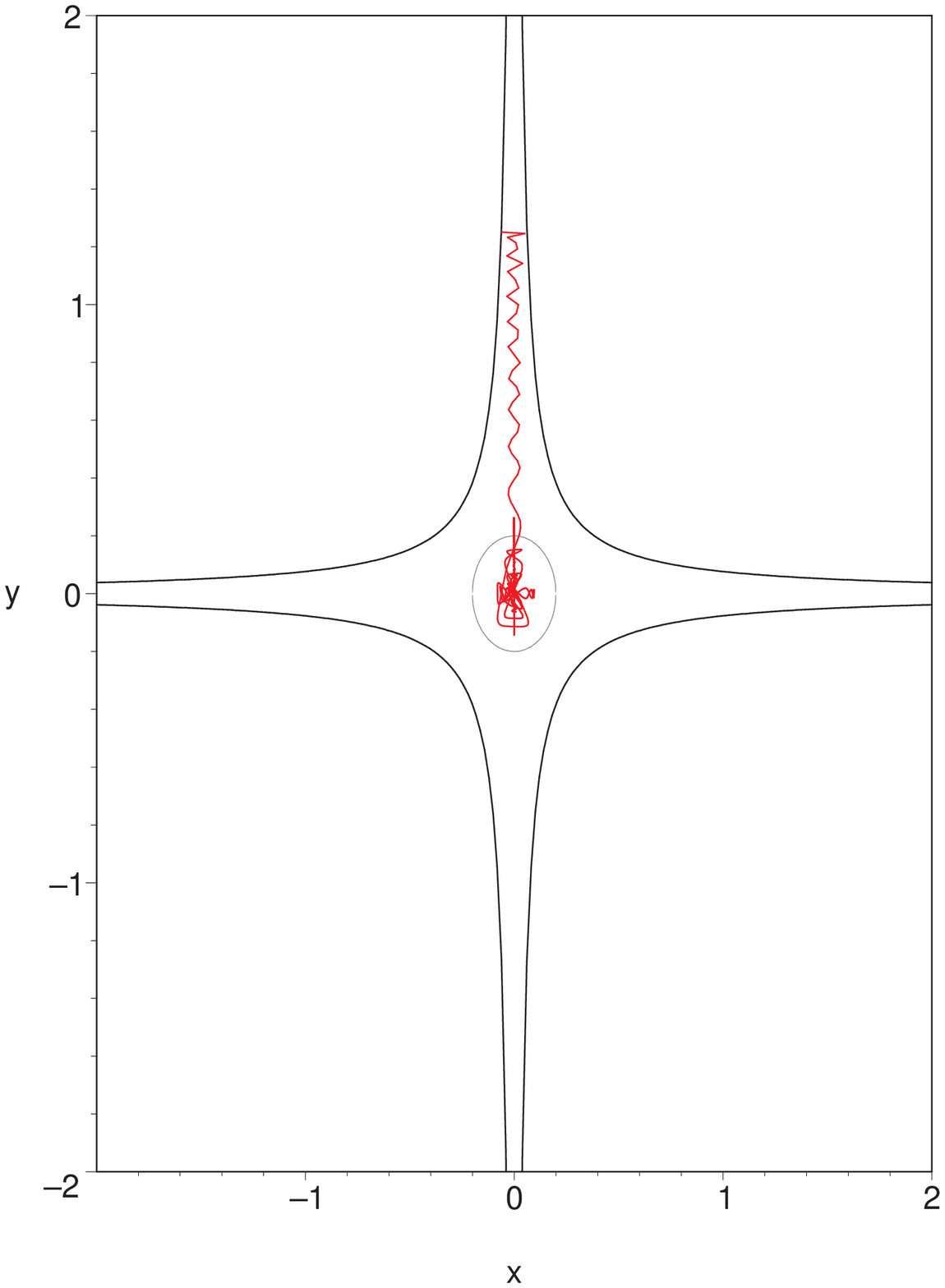}
\end{center}
\epsfxsize=0.4\textwidth
\begin{center}
\leavevmode
\epsffile{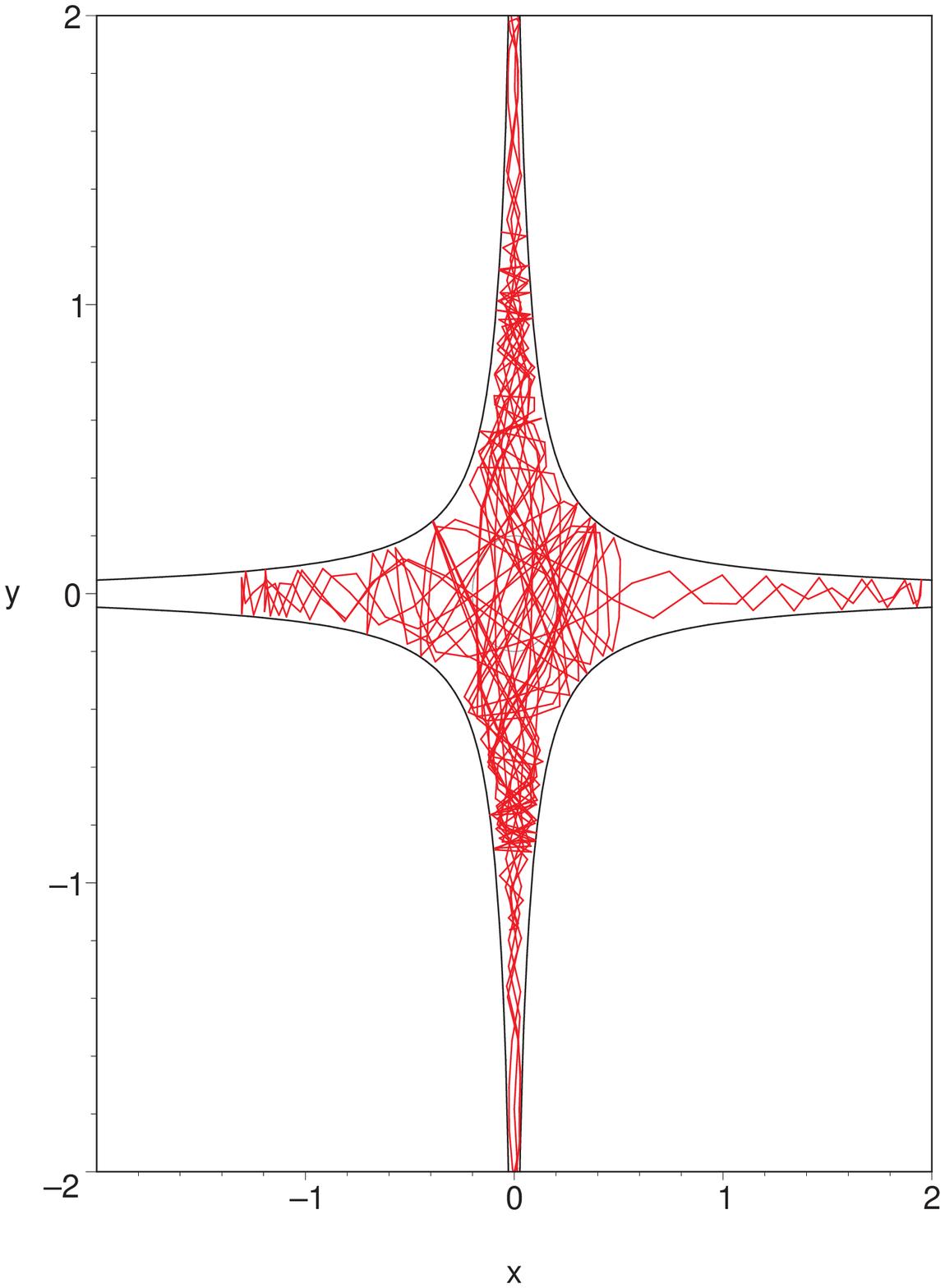} \,  
\epsfxsize=0.4\textwidth
\epsffile{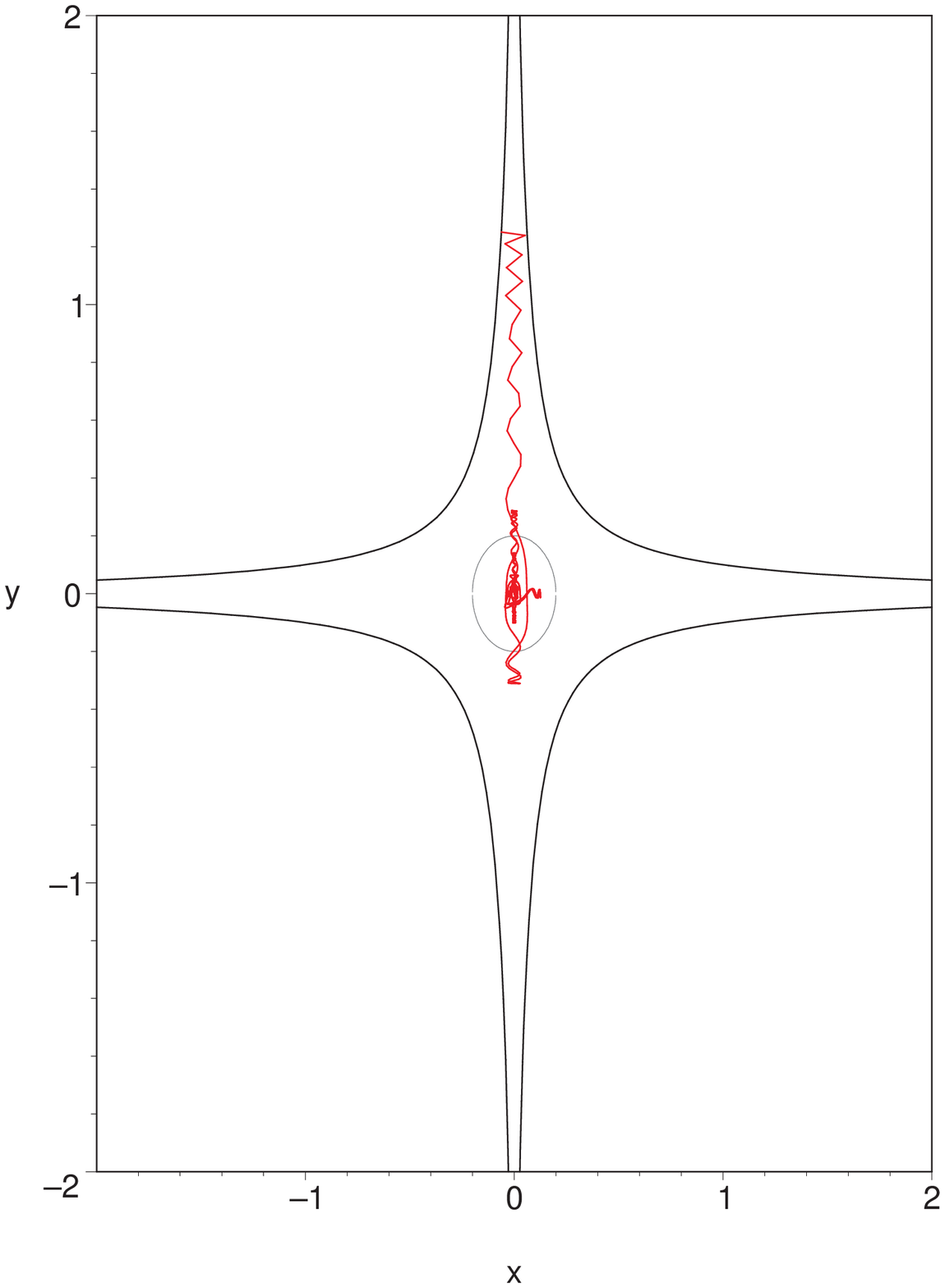}
\end{center}
\parbox[c]{\textwidth}{\caption{\label{vier}{\footnotesize The solutions (\ref{3.2}) arising from the setups i) (upper left), ii) (upper right), iii) (lower left), and iv) (lower right). The black hyperbolic lines are equipotential lines indicating the initial energy in the matter sector while the gray circle around the origin defines the stadium. Neglecting gravitational effects (left column) the solutions sample the allowed field space while including gravitational effects (right column) induces a trapping of the solution in the central region.}}}
\end{figure}
The resulting trajectories in the $(x,y)$-plane are shown in 
figure \ref{vier}.
Here the upper left, upper right, lower left and lower right diagram display
the numerical solutions corresponding to case  i), ii), iii), and iv), 
respectively. The hyperbolic black lines in these diagrams are the 
equipotential lines on which the solutions start with zero 
initial velocity. We have also added
(in gray) a geodesic circle, which indicates the region we take to 
be the stadium.\footnote{Note that we do not have an intrinsic criterion  
which defines the stadium. Nevertheless we think that the choice
we have made is helpful for visualizing the qualitative features 
of the solutions.}

Figure \ref{vier} immediately tells us two important messages.
The first is that there is no difference in the qualitative 
behavior between the full conifold model and its leading order
truncation. In both cases the potential (away from the bottom of the valley) has a gradient pointing in the direction of the central region. This gradient prevents the solutions from running off to infinity and turns them back to the stadium. The qualitative behavior of the solutions then depends on the presence of this property and is independent of the detailed shape of the potential.
%
%
 
The second message is that Hubble friction drastically changes the 
dynamics of the solution. If it is neglected, the scalar fields
sample the region below the equipotential line on which
they start. Although solutions never go very deeply into the 
valleys, and always return to the 
stadium
after a while, they spend a considerable amount of time 
out in the valley. Thus there is a trapping of the moduli, but it is not
very efficient.\footnote{This is the situation described in \cite{beauty}.} 
However, the picture changes 
drastically once Hubble friction is taken into account. Now
we observe that the solutions just travel down the valley to the
stadium and stabilize close to the conifold point. 

This
different behavior is easily understood in terms of 
the energy carried by the solutions. If Hubble friction 
is switched off in (\ref{7.4}), (\ref{7.5}) we have a
geodesic motion modified by a force term, and the total
energy $T+\pV$ is conserved. This is clearly visible in 
figure \ref{vier} where the solutions in the left column 
essentially
sample the whole region which is energetically accessible.
We say `essentially' because the solutions never move
very deeply into the valley although this is not forbidden
by energy conservation.
Once Hubble friction is
included in (\ref{7.4}), (\ref{7.5}), the total energy $T+\pV$
of the scalar fields is no longer conserved. In an (overall) expanding
universe, where $3 \ad + \bd> 0$, $T+\pV$ decreases, and this
damping leads to the immediate trapping of the solutions close to the
conifold point.\footnote{In an (overall) contracting universe,
$3 \ad + \bd <0$, energy is transferred from the gravitational
to the scalar sector and solutions are `driven uphill.' Thus
a cosmological bounce is a natural mechanism for moving the 
moduli from one ESP into another one, which might be
`far away.'}
The fact that Hubble friction is very efficient in damping the
motion of scalar fields in an expanding universe is of course
well known. 
In principle, it can also happen that Hubble friction
stops the motion of the moduli before such an ESP is reached \cite{beauty}. The interesting point concerning our solutions is that (provided the initial values of $\ad$ and $\bd$ are not chosen too large) they generically reach the stadium and get trapped at the conifold point.
%
%
%
This effect is due
to an interplay of two ingredients: the gradient of the scalar 
potential and Hubble friction. Without Hubble friction 
the trapping is inefficient, while Hubble friction alone
only traps the solution at the bottom of the valley and not necessarily at the (physically interesting) ESP.

\end{section}

\begin{section}{Discussion and Conclusions}
In this paper we studied cosmological solutions of M-theory compactified on a
Calabi-Yau threefold in the vicinity of a conifold singularity. The dynamics
is described by a recently constructed low energy effective action  
\cite{CFL,DIS} which explicitly includes the 
M2-brane winding modes that become massless at the conifold point. 
Within this framework we found cosmological solutions which dynamically pass through the conifold transition. These solutions interpolate between Calabi-Yau compactifications with different massless field content and gauge group. Describing such dynamical transitions thereby requires that the dynamics of the winding modes is explicitly taken into account.
While solutions which undergo topological transitions exist, they
are suppressed while solutions which get trapped in the
region of the conifold ESP are generic. Thus we see that the 
transition locus is dynamically preferred, even though the potential
still has flat directions. In our case the moduli trapping 
occurs while supersymmetry is unbroken and the flat directions
of the potential are not lifted.

The underlying effect is an interplay between
the shape of the scalar potential and Hubble friction. The details
of the potential are not relevant, in particular the full fledged
supergravity model for a conifold transition and its non-supersymmetric
first order approximation display the same type of behavior. This is
similar to flop transitions where the scalar potential 
does not have a Higgs branch. Here, the 
non-supersymmetric truncations
\cite{flop} and the full
supergravity model \cite{FS4} have the same qualitative dynamics. 
In both cases one observes perfect trapping, in the sense that the (numerical)
solutions never left the transition region, despite that the
potential had many flat directions. It remains, however, an open question to what extend the trapping depends
on the number of flat directions. In this paper we have 
truncated the dynamics down to two scalars, corresponding to two
one-dimensional branches, while in full string compactifications 
these branches are higher dimensional. Each additional 
flat direction opens a new opportunity for the scalar fields
to escape the stadium and might weaken the trapping effect.\footnote{We thank 
Burt Ovrut for raising
this issue.}

In \cite{beauty} a different effect called  quantum moduli trapping
was discussed,  which also works towards the
stabilization of the moduli at ESPs. Here the trapping mechanism results from 
the quantum production of light particles in the vicinity 
of an ESP. This process extracts energy from the coherent 
motion of the scalar fields and converts it into particles.
The feedback of the particle production on the scalar field
induces an effective potential which drives the scalar field
towards the ESP. The analysis of \cite{beauty} focused on 
fast trapping, meaning trapping at time scales small compared to
the Hubble time, $t \ll H^{-1}$. But it was also argued that 
quantum trapping also happens in an expanding universe. In this case
Hubble friction plays an ambiguous role: it is potentially dangerous, 
because it can stop the scalar motion before an ESP is reached,
but it is also needed to damp
the oscillatory behavior of scalar fields induced
by the quantum potential, so that the scalars really stabilize. 
Thus the mechanism is an interplay between a quantum effective
potential (induced by particle production) with Hubble friction.
A particularly interesting feature of quantum trapping is that 
inflation can arise as  `trapped inflation' \cite{beauty}.

In contrast, the potential featuring in our mechanism is 
`classical,' in the sense that particle production is not taken 
into account.\footnote{Since our action is to be interpreted as
an effective action, it is not quite correct to call it classical.
In fact we have `integrated in' certain non-perturbative
states, namely winding states of M2-branes.} Since we include
Hubble friction, we find that classical trapping is much more
efficient than argued in the Appendix D of \cite{beauty}.
However, the classical trapping 
mechanism cannot produce inflation. In this respect the
dynamics of the conifold potential is similar to the one of
the flop model \cite{FS4}: one can find short periods of
accelerated expansion at collective turning points of the scalars,
but the potential is either flat, but vanishing or non-vanishing,
but too steep to sustain the accelerated expansion for an extended period of time. This situation might be ameliorated by switching on background
fluxes to gently lift some of the flat directions.

Our conclusion is that both mechanisms, classical and quantum moduli trapping, are relevant and important
for string cosmology. It would therefore be very interesting
to include quantum trapping in the model studied in this paper.
Other obvious extensions are the inclusion of flux, 
four-dimensional models or five-dimensional brane-world type models,
and ESPs with non-Abelian gauge groups. This should give the
answer to two important questions: (i) is the dynamics of
string cosmologies such that they are generically attracted to
interesting ESPs, and (ii) can one naturally have inflation 
`on the way' to the ESP? Concerning point (i) it would be
interesting to have an explicit realization of the `enhancement
cascades' mentioned in \cite{beauty}. String moduli spaces 
have a hierarchy of special submanifolds with increasing numbers
of symmetries and light particles for decreasing dimension of the
ESP locus. One therefore expects that the moduli first move
to an ESP hypersurface, and subsequently to subspaces of higher and higher
Co-dimension. It would be interesting to see whether the
resulting enhancement cascades have a high probability to 
end with a realistic particle spectrum, meaning a small
standard model or GUT-like gauge group times a hidden sector. 

\end{section}
\begin{section}*{Acknowledgments}
This work is supported by the DFG within the `Schwerpunktprogramm
String\-theorie'. F.S.\ acknowledges a scholarship from the `Studienstiftung
des deu\-tschen Volkes' and FOM. We thank Burt Ovrut for useful discussions
on the role of Hubble friction and flat directions for moduli stabilization.
\end{section}

\end{document}